\documentclass[twocolumn,aps,pre,showpacs,amsmath,amssymb]{revtex4}

\usepackage{graphicx}
\usepackage{dcolumn}
\usepackage{bm}

\begin{document}
\title{Liquid drop splashing on smooth, rough and textured surfaces}
\author{Lei Xu\footnote{Present address: Harvard University, Cambridge, MA 02138,
USA. Electronic address: xuleixu@deas.harvard.edu}}
\affiliation{
Department of Physics, The University of Chicago, Chicago, Illinois
60637}
\date{\today}
\begin{abstract}
Splashing occurs when a liquid drop hits a dry solid surface at high
velocity.  This paper reports experimental studies of how the splash
depends on the roughness and the texture of the surfaces as well as
the viscosity of the liquid. For smooth surfaces, there is a
``corona'' splash caused by the presence of air surrounding the
drop.  There are several regimes that occur as the velocity and
liquid viscosity are varied. There is also a ``prompt'' splash that
depends on the roughness and texture of the surfaces.  A measurement
of the size distribution of the ejected droplets is sensitive to the
surface roughness. For a textured surface in which pillars are
arranged in a square lattice, experiment shows that the splashing
has a four-fold symmetry.  The splash occurs predominantly along the
diagonal directions. In this geometry, two factors affect splashing
the most: the pillar height and spacing between pillars.
\end{abstract}
\maketitle

\noindent\textbf{I. Introduction}
\\
\\
\indent    When a liquid drop hits a solid surface, it often
splashes and breaks into thousands of smaller droplets.  Splashing
is an excellent example of a singular breakup phenomenon with an
underlying instability that is still not properly understood.  As
illustration, it was only recently discovered that the surrounding
air pressure is an important parameter for creating a splash on a
smooth dry substrate so that the splash can be completely suppressed
in a low pressure environment \cite{Xu2005}.   Splashing is also
broadly important in industry with applications in ink-jet
printing\cite{inkjet}, combustion of liquid fuel\cite{combustion},
spray drying\cite{drying} and surface coating\cite{coating}.

There are two distinct types of splashing\cite{Riboo}: ``corona''
and ``prompt''. Corona splashing occurs on smooth surfaces, where a
symmetric corona is first formed, and droplets are ejected from the
expanding corona; prompt splashing takes place on rough surfaces,
where there is no corona, and droplets are created at the spreading
contact line.  Fig. 1 shows photographs of the two cases.  A
previous study proposed to explain this difference: corona splashing
is caused by the effects of the air surrounding the drop and prompt
splashing is caused by the effects of surface roughness
\cite{arxiv}.

\begin{figure}[!t]
\begin{center}
\includegraphics[width=3.2in]{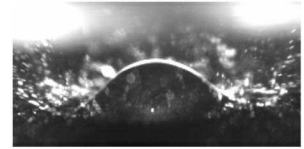}%
\caption{Corona splash and prompt splash. The photograph on the left
is a corona splash on a smooth dry surface. Droplets are created
from a symmetric corona. In the photograph on the right, a prompt
splash occurs on a rough dry surface. In this case there is no
corona and droplets are ejected from the advancing contact line.}
\end{center}
\end{figure}

Splashing has been studied since the time of Worthington in 1867
\cite{Worthington}. Since then, there have been many experimental
studies which have attempted to find a criterion for when splashing
would occur.  Notably, Mundo \emph{et al}\cite{Mundo} established an
empirical relationship for the no-splashing to splashing transition
that depended on the surface roughness, $R_a$, the velocity of
impact, $V_0$, the surface tension of the fluid, $\sigma$, the
diameter of the drop, $D$, the dynamic viscosity of the fluid,
$\mu$, and the density of the fluid, $\rho$. They found that:
$We^{1/2} \cdot Re^{1/4}=K_c[R_a]$, where $We$ and $Re$ are Weber
number and Reynolds number respectively: $We=\rho D V_0^2 / \sigma
$, $Re=\rho D V_0 / \mu$. $K_c$ is a constant and that depends on
surface roughness $R_a$. Splashing will occur when $K>K_c\sim 50$.
When $K<K_c$ no splashing will occur.  Wu\cite{Wu} and Range
\emph{et al}\cite{Range} investigated the dependence of splashing on
the Ohnesorge number, $Oh = \mu / \sqrt{D \sigma \rho}$. They
studied the $Oh << 1$ case , where they could neglect the effects of
viscosity, and obtained the relationship: $We_c=a\cdot log^bR_a$,
where $a$ and $b$ are fitting parameters. When $We > We_c$, they saw
a splash. Note that neither of these relationships take into account
the effects of the gas surrounding the liquid during the splash.

    Some researchers also investigated the fingering instability at the rim of
the expanding liquid disc.  Allen\cite{Allen} proposed that the
Rayleigh-Taylor instability caused the fingering.  Bhola \emph{et
al}\cite{Bhola} and Mehdizadeh \emph{et al}\cite{Mehdizadeh}
obtained reasonable agreement between this theory and their
experiments. Thoroddsen \emph{et al}\cite{Thoroddsen} experimentally
studied the fingers and proposed that the instability is caused by
the presence of air trapped under the liquid drop.

     This paper reports on experiments both for corona splashing on smooth
dry surfaces and for prompt splashing on rough and textured dry
substrates.  For corona splashing, there are several regimes that
depend on the velocity of impact and the fluid viscosity.
Undulations around the rim of the spreading fluid are measured as a
function of the air pressure on the smooth dry surfaces. There is a
sharp jump in the number of undulations at the threshold pressure.
For prompt splashing, both random roughness and roughness created by
regular textured surface were studied.  For a textured surface
consisting of a regular array of pillars, the dependence of
splashing on the vertical pillar height, lateral pillar size and
pillar spacing was studied independently.
\\
\\
\noindent\textbf{II. Corona splash on smooth surface}
\\
\\
\begin{figure}[!b]
\begin{center}
\includegraphics[width=3.2in]{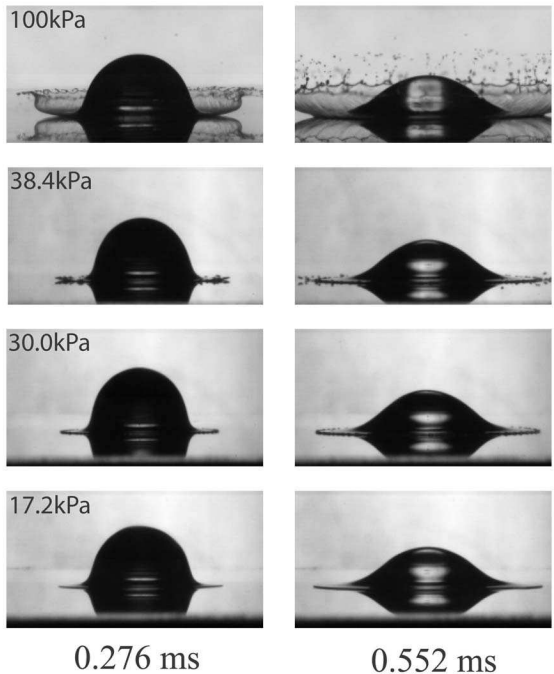}%
\caption{Photographs of a liquid drop hitting a smooth dry
substrate.  A 3.4$\pm$0.1 mm diameter alcohol drop hits a smooth
glass substrate with impact velocity $V_0=3.74\pm0.02$ m/s at
different background air pressures.  Each row shows the drop at two
times, $0.276$ ms and $0.552$ ms after impact. In the top row,
$P=100$ kPa (atmospheric pressure), the drop splashes.  In the
second row, at the threshold pressure, $P_T$ = $38.4$ kPa, the drop
emits only a few droplets, traveling at a small angle with respect
to the surface.  In the third row, at $P = 30.0$ kPa, there is no
splashing but there are undulations at the rim.  In the fourth row,
at $P = 17.2$ kPa, there is no splashing and no apparent undulations
in the rim of the drop. Taken from \cite{Xu2005}.}
\end{center}
\end{figure}
\indent    Previous experiment has shown that the surrounding air is
crucial for corona splashing on a smooth dry surface \cite{Xu2005}:
The rows of Fig. 2 show images of splash at different background air
pressures for a drop of ethanol hitting a glass substrate.
Surprisingly, as the pressure is lowered, fewer droplets are
ejected; under low enough pressure no droplets emerge at all after
impact.  At a threshold pressure, $P_T$, the splash just begins to
be formed as is shown in the second row of the figure.

    The threshold pressure, $P_T$, as a function of impact
velocity, $V_0$, is shown in the main panel of Fig. 3.  The curve is
not monotonic.   In the high velocity region above a characteristic
velocity, $V^*$, $P_T$ decreases as the impact velocity is raised.
This is what we might naively expect.  However, in the region $V_0 <
V^*$, the curve is non-monotonic.  This non-monotonicity indicates
two different regimes at low and high velocities.  Further
experiments show that $V^*$ varies with liquid viscosity and drop
size \cite{unpublished}.

    Experiments have also revealed that when the surrounding gas
is heavier (for example, using Kr and SF$_6$) and with larger liquid
viscosity it is easier to create a splash.  We compared two
stresses\cite{Xu2005}: the destabilizing stress from air,
$\Sigma_G$, and the stabilizing stress from surface tension,
$\Sigma_L$, and found:

\begin{equation}\label{eq1}
\Sigma_G/\Sigma_L = \sqrt{\gamma M_G} P\; \sqrt{\frac{D V_0}
{4k_BT}} \;\, \frac{\sqrt{\nu_L}}{\sigma}
\end{equation}

\begin{figure}[!b]
\begin{center}
\includegraphics[width=3.2in]{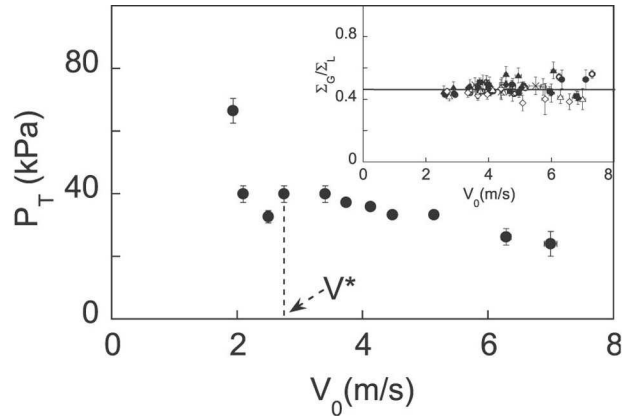}%
\caption{Threshold pressure versus impact velocity and the collapse
of data. Main panel shows $P_T$ vs. $V_0$ in air.  The curve is
nonmonotonic: seemingly two distinct  regimes are separated by a
velocity, $V^*$. The inset plots $\Sigma_G/\Sigma_L$ versus $V_0$ at
threshold pressure $P_T$, in the region $V_0 > V^*$, for gases of
different molecular weight, $M_G$: He($M_{He}$ = 4), air($M_{air}$ =
29), Kr($M_{Kr}$ = 84), SF$_6$($M_{SF_6}$ = 146) and for liquids of
different viscosity, $\nu_L$: Methanol($\nu_{Meth}$ = 0.68cSt),
Ethanol($\nu_{Etoh}$ = 1.36cSt), 2-Propanol($\nu_{2-Pro}$ =
2.60cSt). At threshold pressure, all $\Sigma_G/\Sigma_L$ collapse
approximately onto a constant value, 0.45. Taken from
\cite{Xu2005}.}
\end{center}
\end{figure}

Here $\gamma$ is the adiabatic constant of the gas, $M_G$ is the gas
molecular weight, $k_B$ is Boltzmann's constant, $T$ is the
temperature, $D$ is the diameter of the drop, $\nu_L$ is the
kinematic viscosity of liquid, and $\sigma$ is the surface tension.
A heavier gas or a larger liquid viscosity will increase the ratio,
$\Sigma_G/\Sigma_L$. The ratio of these two stresses was found to be
approximately constant for velocities above $V^*$ at threshold
pressure. This is shown in the inset to Fig. 3 where
$\Sigma_G/\Sigma_L$ at threshold pressure $P_T$ is plotted for gases
of different molecular weights(4 $\sim$ 146 Dalton), liquids of
different viscosities (0.68 $\sim$ 2.6 cSt) and different impact
velocities (2.5 $\sim$ 7 m/s). At threshold pressure, in the regime
$V > V^*$, $\Sigma_G/\Sigma_L=0.45$ so that Eq.1 successfully
collapses all the data without any fitting parameter.

\begin{figure}[!b]
\begin{center}
\includegraphics[width=3.2in]{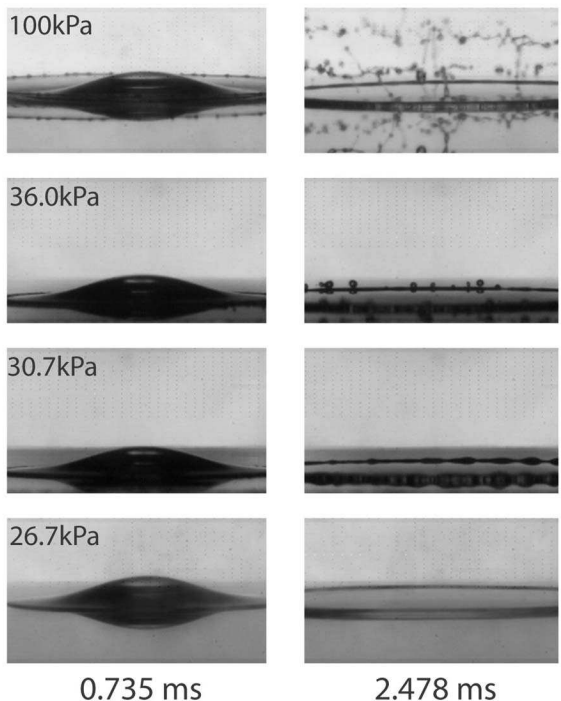}%
\caption{Splash of a viscous drop($\nu_L = 5cSt$).  A $3.1 \pm 0.1$
mm diameter silicone oil drop hits a smooth glass substrate at
impact velocity $V_0=4.03\pm0.05$ m/s under different air pressures.
Each row shows the drop at two times: $0.735$ ms and $2.478$ ms
after impact.  In the top row, at $P = 100$ kPa, there is a
pronounced splash. In the second row, at threshold pressure, P$_T$ =
$36.0$ kPa, the drop just starts to splash.  In the third row, at $P
= 30.7$ kPa, there is no splash but there are undulations in the
thickness of the rim.  In the fourth row, $P = 26.7$ kPa, there is
no splashing and no apparent undulations in the rim.  The general
property that less air leads to less splashing is similar to the low
viscosity case shown in Fig. 2.  However, at high viscosity
splashing occurs at a later time (2.478 ms for the second row) than
it does for low viscosity (0.552 ms for Fig.2 the second row).}
\end{center}
\end{figure}

\begin{figure}[!b]
\begin{center}
\includegraphics[width=3.2in]{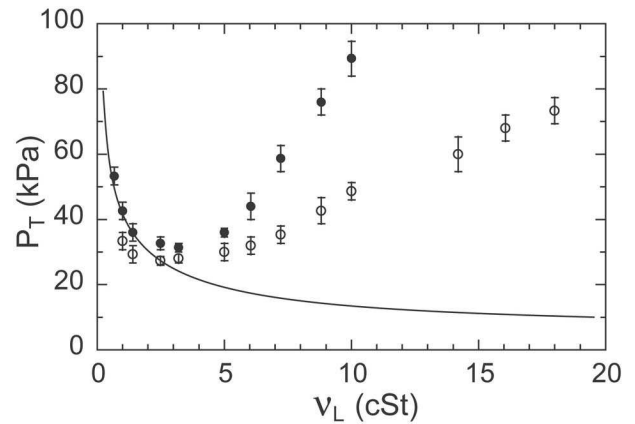}%
\caption{Threshold pressures versus viscosity. Except for the point
at the smallest viscosity, which is for methanol, the fluids were
silicone oils with different molecular weights. Two curves are
shown: (i) the splash threshold pressure, $P_T(\bullet)$, for a
splash to appear and (ii) the bump threshold pressure,
$P_{T-bump}(\circ)$, where the bump at the leading edge first
appears.  The curves are both non-monotonic.  At low viscosity, the
threshold pressures decrease with increasing  $\nu_{L}$ whereas at
high viscosity, they increase with $\nu_{L}$.  The solid line is the
curve predicted by Eq.1: $P_T \sim 1/\sqrt{\nu_{L}}$. The curve fits
the small viscosity regime very well, but does not capture at all
the trend at high $\nu_{L}$.  In these experiments, the impact
velocity and drop diameter are kept fixed at $V_0 = 4.03 \pm 0.05$
m/s, $D = 3.1 \pm 0.1$ mm. }
\end{center}
\end{figure}

    The prediction of Eq.1 that increasing the
liquid viscosity leads to a lowering of the threshold pressure was
verified by the data in Fig. 3 which spanned the range $0.68 cSt <
\nu_L < 2.60 cSt$.  Nevertheless this result is counterintuitive
from our experience with high viscosity liquids and calls for more
experiments covering a broader range of viscosity.  By using
silicone oils of different molecular weights, the liquid viscosity
could be varied by more than one order of magnitude, while keeping a
very similar mass density ($0.82 \sim 0.95 g/cm^3$) and surface
tension($17.4 \sim 21$mN/m).  Fig. 4 shows photographs of a
relatively viscous ($5$ cSt) silicone oil drop hitting a dry glass
substrate under different pressures of air.  Again, we find that the
splash decreases as the air pressure is decreased, and that no
splash occurs when the background pressure is low enough.  But one
difference between Fig.4 and Fig.2 is that splashing occurs at a
much later time when the viscosity is large. This is most obvious if
one compares the second rows. Clearly one effect of viscosity is to
delay the splashing time, as we might have expected.

    Fig.5 shows the threshold pressure, $P_T$, vs. liquid viscosity,
$\nu_L$, for $3.1 \pm 0.1$ mm diameter drops hitting the substrate
with an impact velocity $V_0 = 4.03 \pm 0.05$ m/s. The upper curve
shows, as before, the splashing threshold pressure, $P_T$, where
splashing is first detected.   The lower curve shows the threshold
pressure, $P_{T-bump}$, for where an undulation in the expanding
sheet of liquid is first observed.  $P_{T-bump}$ is defined as the
lowest pressure at which undulations (or bumps) first show up, and
below which no undulations can be seen.  Both threshold pressures
first decrease then increase with increasing viscosity.  This
indicates two different regimes. At low $\nu_L$, as the viscosity is
increased, the threshold pressure to create a splash decreases. Thus
viscosity helps to produce splash, as predicted by Eq.1.  The solid
line is the scaling relation derived from Eq.1, which agrees well
with the small $\nu_L$ data.  However, the prediction starts to
deviate at higher viscosity where the threshold pressures increase
with $\nu_L$.  In this regime, the higher the viscosity, the higher
the pressure of air needed to create a splash and viscosity
suppresses splashing.

    Why there are two different behaviors? We think that for the
low $\nu_L$ regime, the expanding liquid film is stabilized mainly
by surface tension so that viscosity only affects the film
thickness: $d \sim \sqrt{\nu_Lt}$. Thus a larger $\nu_L$ causes a
thicker film which is easier to destabilize.  But for the high
$\nu_L$ regime, viscous drag is important and helps to stabilize the
spreading drop.
\\
\\
\noindent\textbf{III. Number of undulations versus pressure}
\\
\\
\indent    Researchers have extensively studied the fingering
instability that occurs as a splash is created
\cite{Allen,Bhola,Mehdizadeh,Thoroddsen}.  ``Fingers'' mean long
protrusions at the rim of the expanding liquid film.  In our
experiment, we also observed undulations around the rim as shown in
the inset to Fig.6.  Here, a feature similar to ``fingering'' is
observed.  However in this case, the undulations do not extend very
far out from the rim.   In order to prevent possible confusion, we
call them ``undulations'' or  ``bumps''.   Previous studies have
focused on the number of fingers as a function of impact velocity
and surface roughness.  Here we concentrate on the behavior and
number of the bumps as the air pressure is varied.

\begin{figure}[!h]
\begin{center}
\includegraphics[width=3.2in]{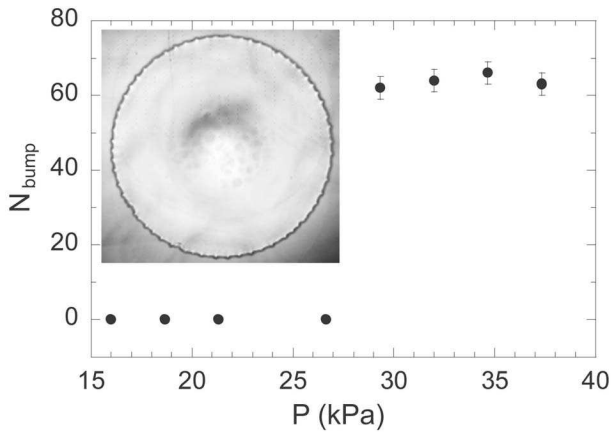}%
\caption{Number of bumps as a function of air pressure. Inset is a
bottom-view photograph showing the undulations.  The number of
undulations shown in the main panel, $N_{bump}$, is counted from
such images.  There is a sudden change in $N_{bump}$ around
$P_{T-bump} = 29$ kPa.   At that pressure $N_{bump}$ jumps from 0
to a finite value and stays constant above that pressure.}
\end{center}
\end{figure}

In Fig.6 we show $N_{bump}$ at different pressures.  We determine
the number of undulations, $N_{bump}$, from pictures such as the one
shown in the inset.  At all pressures, $N_{bump}$ is measured at the
same fixed radius of expansion where the undulations are most clear.
The main panel shows that at threshold pressure $P_{T-bump}$,
$N_{bump}$ jumps from zero to a finite value, and stays constant for
higher pressures. The absence of undulations at low pressure
suggests that no instability can grow below $P_{T-bump}$.  The
apparent pressure independence of $N_{bump}$ above $P_{T-bump}$
could be due to the narrow pressure range we are able to measure:
above
a certain pressure the entire expanding film is lifted into air.\\

\noindent\textbf{IV. Discussion on instability mechanism}
\\

    What is the mechanism for destabilizing the system and
causing the occurrence of a splash on a smooth surface?   What is
the instability that eventually produces a splash? This question is
still under debate. One prevailing theory first proposed by
Allen\cite{Allen}, is that it is due to the Rayleigh-Taylor
instability\cite{Taylor}.  This is an instability occurring at the
interface of two fluids with different densities, when the
acceleration, $a$, points from the light fluid, $\rho_1$, towards
the heavy fluid, $\rho_2$.  Linear instability theory predicts an
exponential growth in amplitude, $A \sim Exp(ct)$, with growth rate
$c$.  The wavenumber, $k_m$, and the growth rate, $c_m$, of the
fastest growing mode are predicted to be\cite{Sharp}:

\begin{equation} \label{eq2}
k_m = \sqrt{\frac{a(\rho_2-\rho_1)}{3\sigma}}
\end{equation}

\begin{equation} \label{eq3}
c_m = \sqrt{\frac{2ak_m(\rho_2-\rho_1)}{3(\rho_2+\rho_1)}}
\end{equation}

Here the light fluid is the surrounding air and the heavy fluid is
the expanding liquid. The expanding disc decelerates so that the
direction of $a$ satisfies the requirement for the Rayleigh-Taylor
instability.  However, since $\rho_2>>\rho_1$, Eqs.2 and 3 do not
vary appreciably with air pressure. Therefore splashing should not
vary with air pressure if it were caused by Rayleigh-Taylor
instability. This is inconsistent with our experiment.

Another interface instability, the Kelvin-Helmholtz instability, can
take place when there is velocity jump at the interface. For
inviscid fluids and $\rho_2>>\rho_1$, the wavenumber and the growth
rate of the fastest growing mode are: \cite{Acheson, Villermaux,
Yoon}:

\begin{equation} \label{eq4}
k_m = \frac{2}{3}\frac{\rho_1 u^2}{\sigma}
\end{equation}

\begin{equation} \label{eq5}
c_m = k_m u \sqrt{\frac{\rho_1}{3\rho_2}}
\end{equation}

With $u$ the relative velocity between two fluids at the interface.
In our case, $u\sim\sqrt{D V_0 / 4 t}$ is the velocity of the
expanding liquid film.  The Kelvin-Helmholtz instability strongly
depends on the density of the lighter fluid, $\rho_1$, and thus may
be relevant to our experiment. However, our previous
results\cite{Xu2005} indicate that compressibility of air is
important.  This suggests that we should replace the Bernoulli term
$\rho_1u^2$ in Eq.4 with $\rho_1 C_Gu$, with $C_G=\sqrt{\gamma
k_BT/M_G}$ the speed of sound in the surrounding gas.

\begin{equation} \label{eq6}
k_m=\frac{2}{3}\frac{\rho_1C_Gu}{\sigma}=\frac{2}{3}\frac{P
}{\sigma}\; \;\sqrt{\frac{\gamma M_G}{k_B T}} \;\,\sqrt{
\frac{DV_0}{4t}}
\end{equation}

The characteristic length in the expanding liquid film is the film
thickness, $d$. This suggests that the instability might be able to
grow if:

\begin{equation} \label{eq7}
k_m\sim 1/d
\end{equation}

Since we also have $d\sim\sqrt{\nu_L t}$, from Eq.6 and Eq.7, we
obtain as a criterion for the instability to grow:

\begin{equation} \label{eq8}
\sqrt{\gamma M_G} P\; \sqrt{\frac{D V_0} {4k_BT}} \;\,
\frac{\sqrt{\nu_L}}{\sigma}\sim 1
\end{equation}

The left hand side is exactly $\Sigma_G/\Sigma_L$ in Eq.1. In the
low viscosity regime, our experiment gives $\Sigma_G/\Sigma_L =
0.45$ for the splashing threshold which is consistent with the
criterion in Eq.8.  This suggests the possibility the
Kelvin-Helmholtz instability may be the underlying instability
mechanism for corona splashing.
\\

\noindent\textbf{V. Prompt splash on rough surface}
\\
\\
\indent A completely different type of splash, the prompt splash,
occurs on rough surfaces. By systematically varying the degree of
surface roughness and the air pressure, we discovered two different
mechanisms for the two kinds of splashes: surrounding air is
responsible for the corona splash discussed above and surface
roughness is responsible for the prompt splash\cite{arxiv}. Under
ordinary conditions (atmospheric pressure and non-zero roughness), a
splash is a mixture of both contributions. By working under low
pressure with a negligible amount of air, we are able to study pure
prompt splashing.

Since a prompt splash is caused by surface roughness, it may retain
information about surface roughness in the distribution of sizes of
the ejected droplets.  We find that this is the case.  We mix a
small amount of ink into our ethanol and then collect the ejected
droplets on a sheet of white paper.  We then obtained the sizes of
ejected droplets by measuring the size and darkness of the stains
left on the paper. Our previous study\cite{arxiv} shows that the
number of droplets, $N$, decays exponentially with their radius,
$r$: $N(r)\sim Exp(-r/r_0)$(see inset of Fig. 9(b)). This indicates
the existence of a characteristic decay length, $r_0$.

The decay length, $r_0$, correlates with roughness of the surface,
$R_a$\cite{arxiv}.  At small $R_a$, we have the relationship
$r_0\approx R_a$; but for large roughness, this breaks down as $r_0$
saturates at a constant value. We can understand this behavior in
the following manner.  After impact, the thickness of expanding
film, $d$, grows continuously from being molecularly thin just after
the impact, to approximately $40 \sim 50 \mu m$ at the end of the
film expansion. When the surface roughness is small, $d$ can grow to
be much larger than $R_a$.  At the beginning, when $d$ is small, the
film is thinner than $R_a$ and continues to eject droplets until the
film becomes much thicker than the surface roughness. After $d$
grows to be larger than $R_a$, the roughness is too small to
destabilize the liquid film and produce a splash.  Thus the
distribution of ejected droplets reflects the surface roughness
$R_a$ and we find $r_0\approx R_a$.  However, when the roughness is
large, $d$ can never grow to be greater than $R_a$. Consequently,
$r_0$ can only grow to the maximum size of $d$ at its final
thickness.  This is consistent with the decay constant, $r_0$,
saturating around $40\mu m$, which is roughly the film thickness at
the end of expansion.\\

\noindent\textbf{VI. Prompt splash on textured surface}
\\
\\
\begin{figure}[!b]
\begin{center}
\includegraphics[width=3.2in]{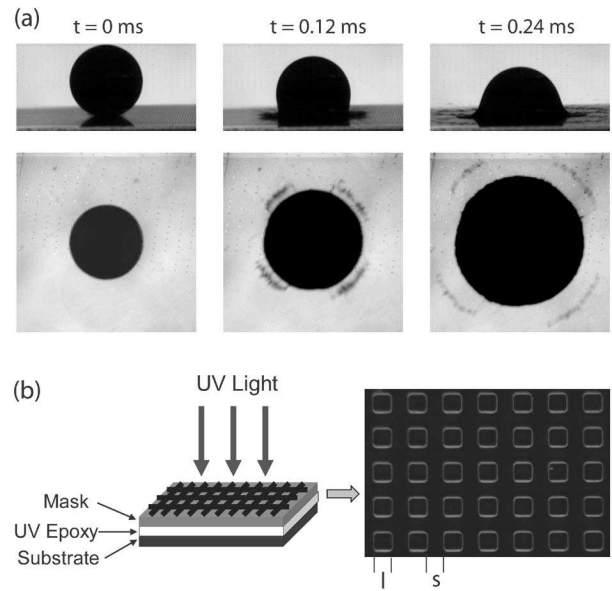}%
\caption{Prompt splashing on a textured surface. (a)Top row shows a
side view of prompt splashing on a textured surface.  Bottom row
shows a bottom view of prompt splashing on a textured surface. There
is a clear four-fold symmetry in the splash which is predominantly
in the diagonal directions of the square lattice created by the
pillars. (b) Making a textured surface with UV-lithography
technique.  Left cartoon shows the UV-lithography process. The right
picture shows a typical textured surface under the microscope. We
define the pillar height as $h$, lateral pillar size as $l$ and
spacing between pillars as $s$. For this particular substrate, $h =
18 \mu m$, $l = s = 60 \mu m$.}
\end{center}
\end{figure}
\indent The last section showed that roughness has a strong effect
on prompt splash. To understand this dependence in more detail and
to understand how surface properties affect splashing, we study
splashing on a well-defined textured surface of regular patterns.

The textured surface is made with UV-lithography: We first spin coat
UV epoxy (SU8-2000, MicroChem Corp.) onto a clean glass microscope
slide.  We then cover the slide with a mask of pre-designed pattern
(square blocks in a square lattice) and expose the slide to UV
light.  After development, the UV epoxy film which is directly under
the transparent part of the mask will harden and the rest of it can
be rinsed away, resulting in a structure on the substrate of square
pillars arranged in a 2D square lattice as shown in Fig. 7b. There
are three important quantities relating to our splash experiments in
this textured surface: (1) the vertical pillar height, $h$, (2) the
lateral pillar size, $l$, and (3) the lateral spacing between
pillars, $s$.  By changing the spin speed, we can vary pillar
height, $h$; by designing different mask patterns, we can vary both
$l$ and $s$ independently. Thus we can vary every aspect of the
structure.

Fig. 7(a) shows photographs of a prompt splash on a textured surface
under low air pressure. The impact velocity is $4.3\pm 0.1m/s$, and
the drop diameter is $D=3.4\pm 0.1mm$. The top row shows a side view
of the splash.  It has a similar look as the prompt splash on an
ordinary rough surface. However, the bottom view shown in the second
row reveals a very striking feature: the splashing occurs with four
fold symmetry.  The droplets are ejected predominantly along the
diagonal directions of the square lattice. Fig. 7(b) shows the
process of UV-lithography and a picture of textured surface under a
microscope.

    We can now vary the profiles of the surface and determine their
effect on the splash and the ejected droplet distribution.  Again we
use the ink spot technique to measure the size distribution of
droplets, as mentioned in last section.  In the first set of
experiments we first keep lateral size constant at $l=s=60 \mu m$,
and vary the vertical height of the pillars, $h$. Fig. 8(a) shows
the number of droplets, $N$, versus their radius, $r$. Similar to
the case with random roughness, we find an exponential decay at
large $r$, $N(r)\sim exp(-r/r_0)$, with a characteristic decay
length, $r_0$, that varies with $h$. Fig. 8(b) shows that $r_0$
varies with $h$ in a nonmonotonic manner. For small $h$, $r_0$
increases with, and has a value comparable to, $h$.  This indicates
that $r_0$ is determined by $h$. However, when $h$ is greater than
$18 \mu m$, the opposite trend occurs: $r_0$ decreases as $h$
increases.  Fig.8(c) shows the sum of the areas created by all of
the ink spots, $A_{tot}$, as a function of $h$. $A_{tot}$ is a
quantity that indicates the total amount of ejected droplets.
Fig.8(b) and (c) have the same shape, indicating that $r_0$ and
$A_{tot}$ are strongly correlated. The decreasing trend for large
$h$ implies that larger roughness leads to less splashing. When $h$
is greater than $60 \mu m$, there is no splash at all.

\begin{figure}[!th]
\begin{center}
\includegraphics[width=3.2in]{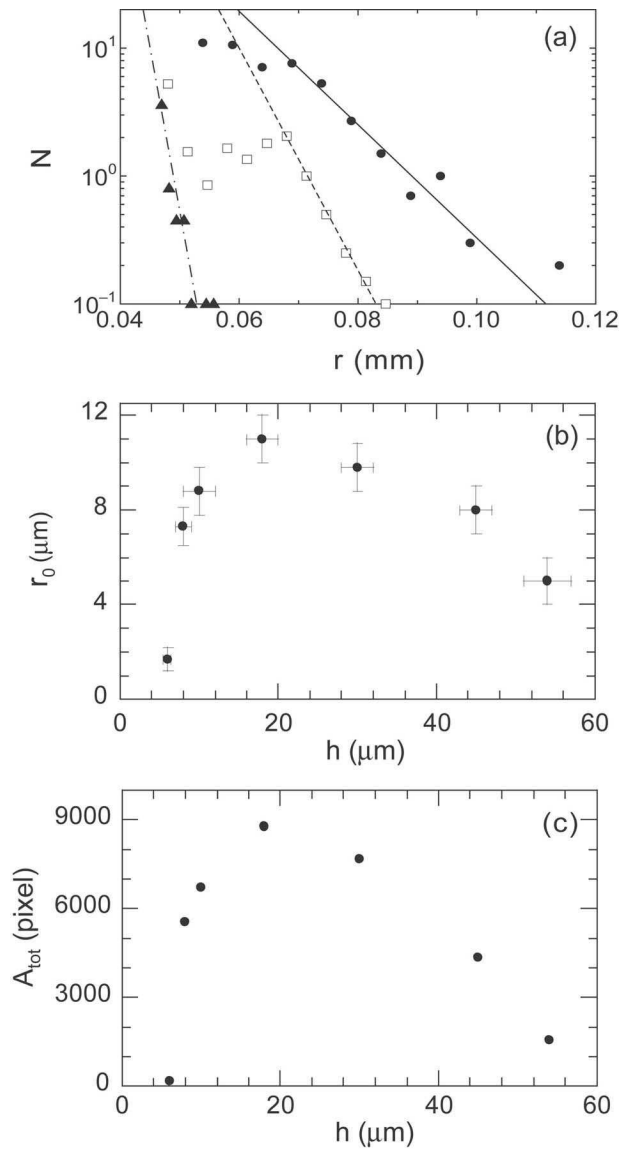}%
\caption{Decay length and total splash area versus pillar height.
(a) The number of ejected droplets, $N$, as a function of droplet
radius, $r$, for three pillar heights, $h$. The lateral dimensions
are kept fixed at $l = s = 60 \mu m$. The straight lines are an
exponential fit to the large $r$ tail of the distribution: $N \sim
Exp(-r/r_0)$, with $r_0$ = $0.0017 mm$  ({--}$\cdot${--}), $0.0098
mm$ ({\large---}) and $0.005 mm$ ({\large{-}{-}{-}}) for h= 6$\mu
m$($\blacktriangle$), h = 30$\mu m$($\bullet$), and h = 54$\mu
m$($\square$) respectively. Note that $r_0$ does not vary
monotonically with $h$. (b) The exponential decay length, $r_0$,
versus pillar height, $h$. $r_0$ first increases then decreases as
$h$ increases. (c) The total ink spot area, $A_{tot}$, as a function
of $h$, for the same set of experiments shown in (b). The curve in
(c) has a similar shape as (b) indicating that $A_{tot}$ and $r_0$
are strongly correlated.}
\end{center}
\end{figure}

    These results are counterintuitive.  We suspect they are caused by
the way in which the impacting liquid drop can flow between the
channels set up by the pillar structure.  At small pillar heights,
the liquid in the drop can easily reach the bottom of the canyon
between the pillars and can then expand along the bottom surface.
During expansion, the liquid film is destabilized by the pillars,
producing droplets with a size related to the pillar height, $h$.
This produces a positive correlation between $r_0$ and $h$ for $h <
18 \mu m$. However, as $h$ increases, it is increasingly difficult
for the impacting drop to reach the bottom surface. In this case, we
suspect that much of the drop expands on top of the pillars, rather
than between them. Once $h \approx 60 \mu m$, that is when the
height is about the same size as the lateral dimensions $l$ and $s$,
the situation resembles a drop expanding on a flat surface with many
holes rather than pillars. Here all the obstructions are underneath
the liquid film and make only a small perturbation to its expansion.
Because the impacting drop can only penetrate a finite depth below
the pillar top, it does not know how far away it is from the bottom
surface. This suggests that the amount of splashing should saturate
as the pillar height is increased. We do not have a good explanation
of the surprising fact that the splashing can be completely
eliminated if the pillars are sufficiently tall.

To some extent, this is similar to the Cassie state of a drop on a
superhydrophobic rough surface studied by Qu\'{e}r\'{e} \emph{et
al}{\cite{David1, David2}}, where a water drop can sit on top of air
trapped in the rough profile of the substrate. However we note that
their case is static whereas ours is probably driven by the fast
dynamics of the expanding drop. Moreover in the case studied by
Qu\'{e}r\'{e} \emph{et al} the air plays an important role in
supporting the weight of the drop. In our situation air has been
pumped out of the system.

\begin{figure}[!t]
\begin{center}
\includegraphics[width=3.2in]{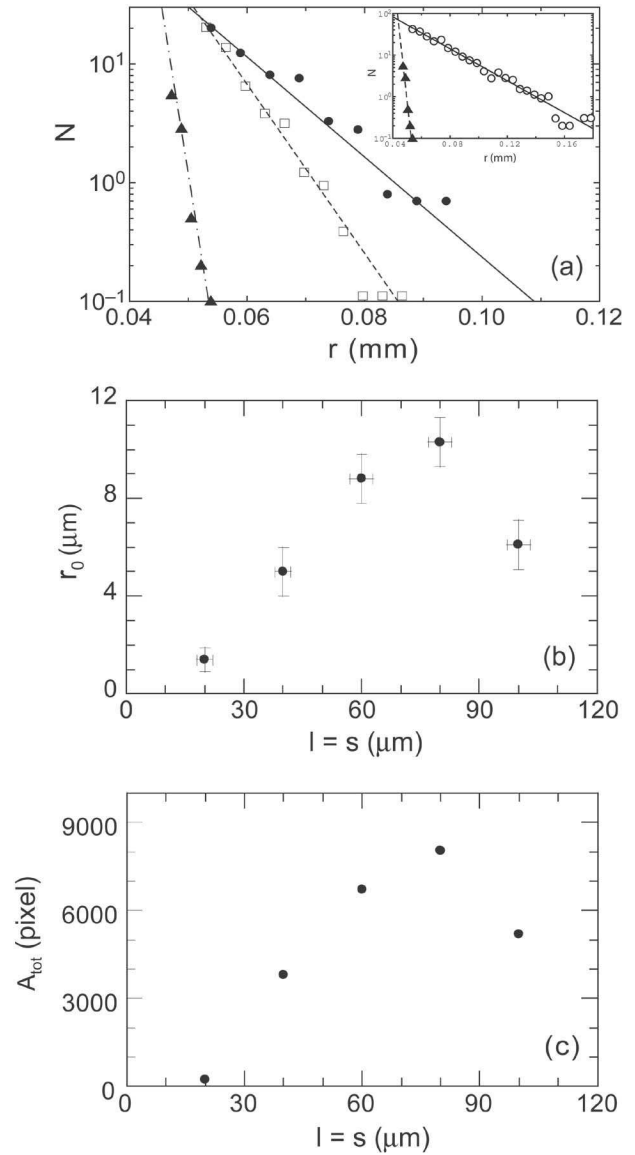}%
\caption{Decay length and total splash area versus lateral size of
the pillars. (a) $N$ as a function of $r$ for three lateral sizes.
The pillar height $h$ is fixed at $h =  10 \mu m$. The straight
lines are an exponential fit to the large $r$ tail of the
distribution: $N \sim Exp(-r/r_0)$, with $r_0$ = $0.0014 mm$
({--}$\cdot${--}), $0.010 mm$ ({\large---}) and $0.0061 mm$
({\large{-}{-}{-}}) for $l = s = 20 \mu m$($\blacktriangle$), $l = s
= 80 \mu m$($\bullet$), and $l = s = 100 \mu m$($\square$)
respectively. Note that $r_0$ does not vary monotonically with $l$
and $s$. Inset compares the distribution for a textured surface with
$l = s = 20 \mu m, h = 10 \mu m$($\blacktriangle$) with a sample
with comparable random roughness $R_a = 16 \mu m(\circ)$. The random
roughness creates a much larger splash. (b) $r_0$ is plotted versus
$l$ and $s$. $r_0$ first increases then decreases with $l$ and $s$.
(c) $A_{tot}$ versus $l$ and $s$. Again we see a similar shape as in
(b).}
\end{center}
\end{figure}

To understand the effect of the lateral dimension on the splashing,
we make substrates of different $l$ and $s$, while keeping $h$
fixed. Fig. 9(a) main panel shows $N$ versus $r$ with an exponential
fitting function, $N \sim Exp(-r/r_0)$, for different lateral sizes.
Fig. 9(b) plots $r_0$ as a function of lateral pillar size $l$ and
spacing $s$.  As $l$ and $s$ are varied we keep $l = s$, and $h =
10\mu m$. Fig. 9(c) shows the total area, $A_{tot}$, vs. $l$ and
$s$. $A_{tot}$ has the same dependence on $l$ and $s$ as does $r_0$.
Both quantities increase with lateral size in most of our range,
then decrease at the end. This means that increasing the lateral
dimensions will enhance splashing for small $l$ and $s$. When the
pillars are too sparse, splashing becomes less pronounced suggesting
that it is more difficult to destabilize the liquid film. We should
also note that $r_0$ is always much smaller than $l$ and $s$, while
much closer to the pillar height $h = 10 \mu m$.  This indicates
that $h$ is more important in determining $r_0$ than are $l$ and
$s$.

A comparison of a textured surface with a random roughness surface
is shown in the inset of Fig.9(a). Both curves decay exponentially,
but random roughness curve (upper curve) has a much larger $r_0$.
This is surprising because both curves have similar roughness ($l =
s = 20 \mu m, h = 10 \mu m$ for the textured surface and $R_a = 16
\mu m$ for the case of random roughness). We can understand this
qualitatively using the data shown in Fig.8(b). There we see $r_0$
starts to decrease at $h\approx \frac{1}{3}l$. Thus here at $h = 10
\mu m = \frac{1}{2}l$, it is already difficult for the drop to reach
the bottom of the substrate and get destabilized. However, the
random roughness is made by particles coated on surface. This
substrate never resembles a flat surface with many holes. Therefore
the random roughness surface makes a much larger splash with a
larger $r_0$.

Fig. 9 demonstrates that $r_0$ changes as we vary $l$ and $s$
together. One further question is whether this is caused by a change
in $l$ or a change in $s$ or in both? We can check this by varying
$l$ and $s$ independently. Fig. 10 shows the result. The different
symbols are for varying $l$ and $s$ separately while leaving all
other conditions unchanged.  Apparently increasing the spacing
between pillars, $s$, enhances $r_0$ and $A_{tot}$ while $l$ has a
much smaller effect on the splashing behavior.

The fact that increasing $s$ enhances splashing helps to explain why
we see splashing in the diagonal directions in Fig. 7a.  Along the
diagonal, the distance between pillars is the greatest.  Because $s$
is largest in those directions, splashing preferentially occurs in
those directions.

\begin{figure}[!t]
\begin{center}
\includegraphics[width=3.2in]{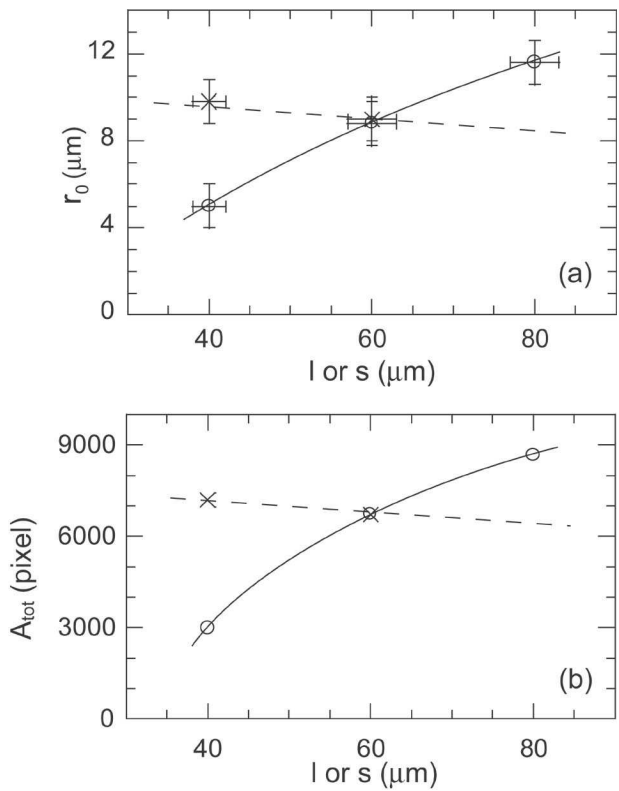}%
\caption{Effect on splash characteristics of varying $l$ as compared
to the effect of varying $s$. (a) We independently vary lateral
pillar size , $l(\times)$, and spacing between pillars, $s(\circ)$,
to compare their effect. Two lines are guides to the eye for $l$
({\large{-}{-}{-}}) and $s$ ({\large---}). Pillar height is kept
fixed at $h = 10 \mu m$. When $l$ is varied, $s = 60 \mu m$ is held
fixed; when $s$ is varied, $l = 60 \mu m$ is held fixed. The
comparison shows that $r_0$ changes with $s$ but not $l$. (b) A plot
of the total area, $A_{tot}$, as a function of $l$ or $s$, for the
same experiment as in (a). It produces the same trend as in (a).}
\end{center}
\end{figure}

The textured substrate not only affects the prompt splashing caused
by surface roughness, it also changes the behavior of corona
splashing caused by the surrounding air. Fig. 11(a) shows a typical
corona splash on a smooth surface at atmospheric pressure, while
Fig. 11(b) shows, at the same pressure, no splash at all on a
textured surface consisting of tall pillars. In both experiments the
drop hits the substrate at the same impact velocity.  Different
amounts of splashing can also be achieved by creating pillars with
intermediate heights. These results suggest that the pillars form
channels through which the air can escape so that the importance of
the air for creating the splash is minimized. This discovery
demonstrates another way in which one can suppress splashing.
Moreover, it has the advantage that this suppression can be achieved
without decreasing the gas pressure.
\\
\begin{figure}[!t]
\begin{center}
\includegraphics[width=3.2in]{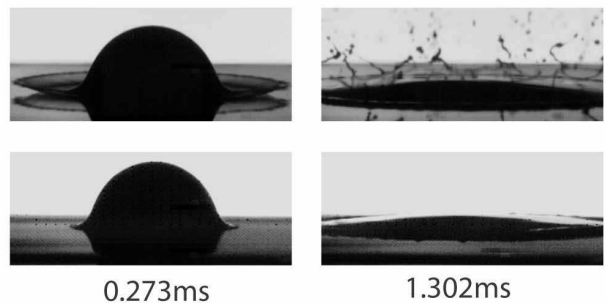}%
\caption{Effect of pillars on suppressing corona splashing. All
experiments are done at atmospheric pressure, with $V_0 = 4.3 m/s$
and $D = 3.4 mm$. (a) A corona splash on a smooth surface. (b)
Splashing is completely suppressed on a  surface consisting of
pillars with $l = s = 60 \mu m$ and $h = 125 \mu m$. We can tune the
amount of splashing by varying the pillar height even under
atmospheric pressure.}
\end{center}
\end{figure}

\noindent\textbf{VII. Conclusion}
\\

\indent This paper systematically studied the splashing of liquid
drops on various dry solid surfaces. This study corroborates that
there are two mechanisms corresponding to the two kinds of splashes.
Air causes the corona splash on smooth dry surfaces and substrate
roughness causes the prompt splash. For the corona splash, we
discovered several regimes.  At high impact velocity, there are two
regimes as the viscosity of the liquid is varied.   We also studied
the finger instability as function of air pressure and find a jump
in the number of bumps. We suspect that Kelvin-Helmholtz instability
coupled with the compressibility of air, is a possible mechanism for
the splashing instability.  This mechanism agrees well with our
experimental data.

   In order to examine the effect of surface roughness, we studied
splashing on textured surfaces consisting of square pillars arranged
in a square lattice.  We found that the dimensions of the pillars
strongly affect splashing. Here the pillar height, $h$, is found to
be the most important factor determining characteristic decay
length, $r_0$.  We discovered that the splash preserves the symmetry
of substrate.  This shows that the splash direction can be
controlled. We also find that corona splash under atmospheric
pressure can be suppressed by making tall pillars on surface. This
provide another way to reduce splash even under normal pressure.
Since splashing is involved in many industrial
processes\cite{inkjet,combustion,drying,coating}, these discoveries
could have important practical applications.

\textbf{Acknowledgement} The author is particularly indebted to
Sidney R. Nagel and Wendy W. Zhang for their suggestions and help on
this work. L.X. is also grateful to Qiti Guo, Jingshi Hu, David
Qu\'{e}r\'{e}, Mathilde Callies-Reyssat and Ling-Nan Zou for helpful
discussions. This work was supported by MRSEC DMR-0213745 and NSF
DMR-0352777. L.X. was supported by Grainger Fellowship.

\end{document}